\documentclass{iaus}
\usepackage{graphicx}
\title[The Magellanic Group and the Seven Dwarfs] 
{The Magellanic Group and the Seven Dwarfs}
\author[Elena D'Onghia \& George Lake]   
{Elena D'Onghia$^{1,2
\thanks{Marie Curie fellow}}$,
 \and George Lake$^2$}

\affiliation{ 
$^2$ Institute for Theoretical Physik, 
 University of Zurich, \\
Winterthurerstrasse 190, 8057 Zurich, Switzerland \\email:{\tt
  elena@physik.unizh.ch; lake@physik.unizh.ch}}

\pubyear{2008}
\volume{256}  
\pagerange{119--126}
\jname{The Magellanic System: Stars, Gas, and Galaxies}
\editors{Jacco Th. van Loon \& Joana M. Oliveira, eds.}
\begin{document}

\maketitle

\begin{abstract}

The Magellanic Clouds were the
largest members of a group of dwarf galaxies that entered the Milky
Way (MW) halo at late times.  This group, dominated by the LMC,
contained $\sim 4$\% of the mass of the Milky Way prior to its
accretion and tidal disruption, but $\approx 70\%$ of the known dwarfs
orbiting the MW.  Our theory addresses many outstanding problems in
galaxy formation associated with dwarf galaxies.  First, it can
explain the planar orbital configuration populated by some dSphs in
the MW.  Second, it provides a mechanism for lighting up a subset of
dwarf galaxies to reproduce the cumulative circular velocity
distribution of the satellites in the MW.  Finally, our model predicts
that most dwarfs will be found in association with other dwarfs.  The
recent discovery of Leo V (Belokurov et al. 2008), a dwarf spheroidal
companion of Leo IV, and the nearby dwarf associations supports our
hypothesis.

\end{abstract}

\firstsection 
\section{Introduction}
 
In the cold dark matter (CDM) model, the dark halos of galaxies like
the Milky Way build up hierarchically, through the accretion of less
massive halos.  When these sub-systems avoid complete tidal
disruption, they can survive in the form of satellite dwarf galaxies.
However, the dwarf galaxies in the Local Group exhibit several
puzzling features.  Numerical simulations of CDM predict 10 to 30
times more satellites within 500 kpc of the Milky Way and M31 than the
modest observed population (e.g. Moore et al. 1999).  This discrepancy
between the expected and known numbers of dwarf galaxies has become
known as the {\it missing dwarf problem}.  The newly discovered
population of ultra-faint dwarfs around the Milky Way and M31 found in
the Sloan Digital Sky Survey increases by a factor of two the number
of known satellites (Simon \& Geha 2007), but goes to even lower
circular velocities where a comparable or even greater increase in the
number of satellites is expected.

Another peculiarity is that many dwarf galaxies in the Local Group lie
in the orbital plane of the Magellanic Clouds and Stream.  These
dwarfs have been associated with the Magellanic Clouds and termed the
Magellanic Group (Lynden-Bell 1976, Fusi Pecci et al. 1995; Kroupa
et al. 2005).  In order to reproduce this planar configuration in the
current scenario for structure formation, Libeskind et al. (2005)
proposed that subhalos are anisotropically distributed in cosmological
CDM simulations and that the most massive satellites tend to be
aligned with filaments.  Similarly, Zentner et al. (2005) suggested
that the accretion of satellites along filaments in a triaxial
potential leads to an anisotropic distribution of satellites.

Systems anisotropically distributed falling into the Galactic halo may
not lie in a plane consistent with the orbital and spatial
distribution of the MW satellites.  For example, a theoretical
bootstrap analysis of the spatial distribution of CDM satellites
(taken from a set of CDM simulations) by Metz et al. (2008) finds that
even if they are aligned along filaments, they will be consistent with
being drawn randomly.  This could mean that alignment of the
satellites along filaments may not be sufficient to reproduce the
observed planar structures.  

As we propose here, the origin of planar distributions is facilitated
by concentrating infalling satellites into groups.

Another issue is that the dSphs of the Local Group tend to cluster
tightly around the giant spirals.  Proximity to a large central galaxy
might prevent dwarf irregulars from accreting material, turning off
star formation, and they may then undergo tidal interactions to
convert them into dwarf spheroidals.  However, isolated dSphs like
Tucana or Cetus found in the outskirts of the Local Group (Grebel et
al. 2003) suggest that dSphs might also form at great distances from
giant spirals prior to their being accreted.  Clues to the questions
raised by these observations may be contained in measurements of the
metallicities of a large sample of stars in four nearby dwarf
spheroidal galaxies: Sculptor, Sextans, Fornax, and Carina.  
Work by
Helmi et al. (2006) shows that all four lack stars with low
metallicity, implying that their metallicity distribution differs
significantly from that of the Galactic halo, indicating a non-local
origin for these systems.  

\section{Why Do Magellanic Clouds need to be accreted in groups of dwarfs?}

We propose that the Magellanic Clouds and seven of the eleven dwarf
galaxies around the MW were accreted as a group that was then
disrupted in the halo of our Galaxy.  This is supported by
observations indicating that dwarfs are often found in associations
and by numerical simulations where subhalos are often accreted in
small groups (e.g. Li \& Helmi 2008).  In particular, the LMC, SMC,
and those dwarfs whose orbits are similar to those of the Magellanic
Clouds may all have originally been part of such a group.  This ``LMC
group'' was dominated by the LMC and had a parent halo circular
velocity of $\sim$75 km s$^{-1}$ with its brightest satellite, the
SMC, having a rotation velocity of $\sim 60$ km s$^{-1}$ as estimated
from its HI distribution. 

There is considerable evidence for tidal debris from the LMC group,
supporting the proposal that it was tidally disrupted.  The LMC and
SMC have been modeled as a pair owing to their spatial proximity; as
either a currently bound pair or one that became unbound on the last
perigalacticon passage.  The number of dwarfs assigned to the
Magellanic Plane Group (Kunkel \& Demers 1976) includes the following
{\it candidates}: Sagittarius, Ursa Minor, Draco, Sextans and LeoII.
Of the dwarfs known before the recent flurry of discoveries, 7 out of
10 within $\sim 200$ kpc might well be part of this group.  The
remaining three -- Fornax, Sculptor and Carina -- have been proposed
to be part of a second grouping (Lynden-Bell 1982).

\section{Evidence for Nearby Associations of Dwarfs}

CDM theory predicts that many dwarf galaxies should exist in the
field.  Numerical simulations show that the normalized mass function
of subhalos is nearly scale-free.  That is, when the circular velocity
distribution function of the subhalos is normalized to the parent
halo, it is nearly independent of the mass of the parent.  Thus,
groups of dwarf galaxies are a natural expectation of CDM models on
small mass scales.  However, like low mass satellites, these systems
are difficult to observe.

\begin{figure}[b]
\begin{center}
 \includegraphics[width=3.4in]{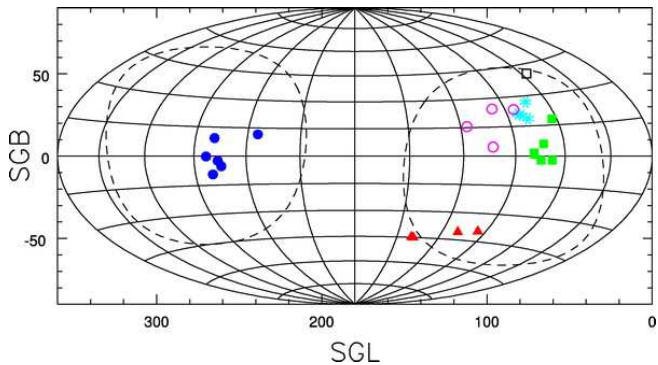} 
 \caption{Distribution in supergalactic coordinates 
of associations of dwarfs galaxies with accurately known 
distances between 1.1 and 3.2 Mpc
(Tully et al. 2006).}
   \label{fig1}
\end{center}
\end{figure}

Tully \etal (2006) discovered a number of associations of dwarf
galaxies within 5 Mpc of the MW. Figure 1 displays the distribution in
supergalactic coordinates of these associations with accurately known
distances between 1.1 and 3.2 Mpc.  These groups have properties
expected for bound systems with 1-10x10$^{11}$ M$_{\odot}$, but are
not dense enough to have virialized, and have little gas and few
stars.  Of the eight associations compiled by Tully (2006), there are
only three for which the two brightest galaxies differ by at least 1.5
magnitudes: NGC3109, NGC1313 and NGC4214.  In the other five, the two
brightest galaxies are certain to merge if the associations collapse
and virialize.

\begin{figure}[b]
\begin{center}
 \includegraphics[width=3.4in]{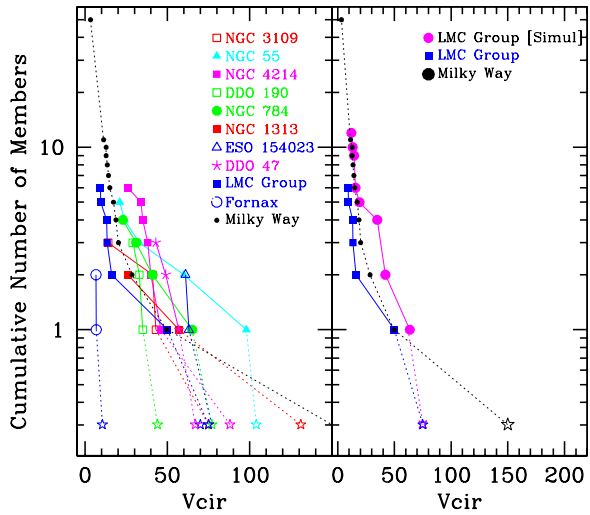} 
 \caption{Cumulative circular velocity distribution of the satellites of the LMC group as compared to the 
nearby dwarf associations (left panel) and to the simulated LMC group in a
$\Lambda$CDM model (right panel) (see D'Onghia \& Lake (2008) for details).}
   \label{fig4}
\end{center}
\end{figure}

\begin{figure}[h]
\begin{center}
 \includegraphics[width=3.4in]{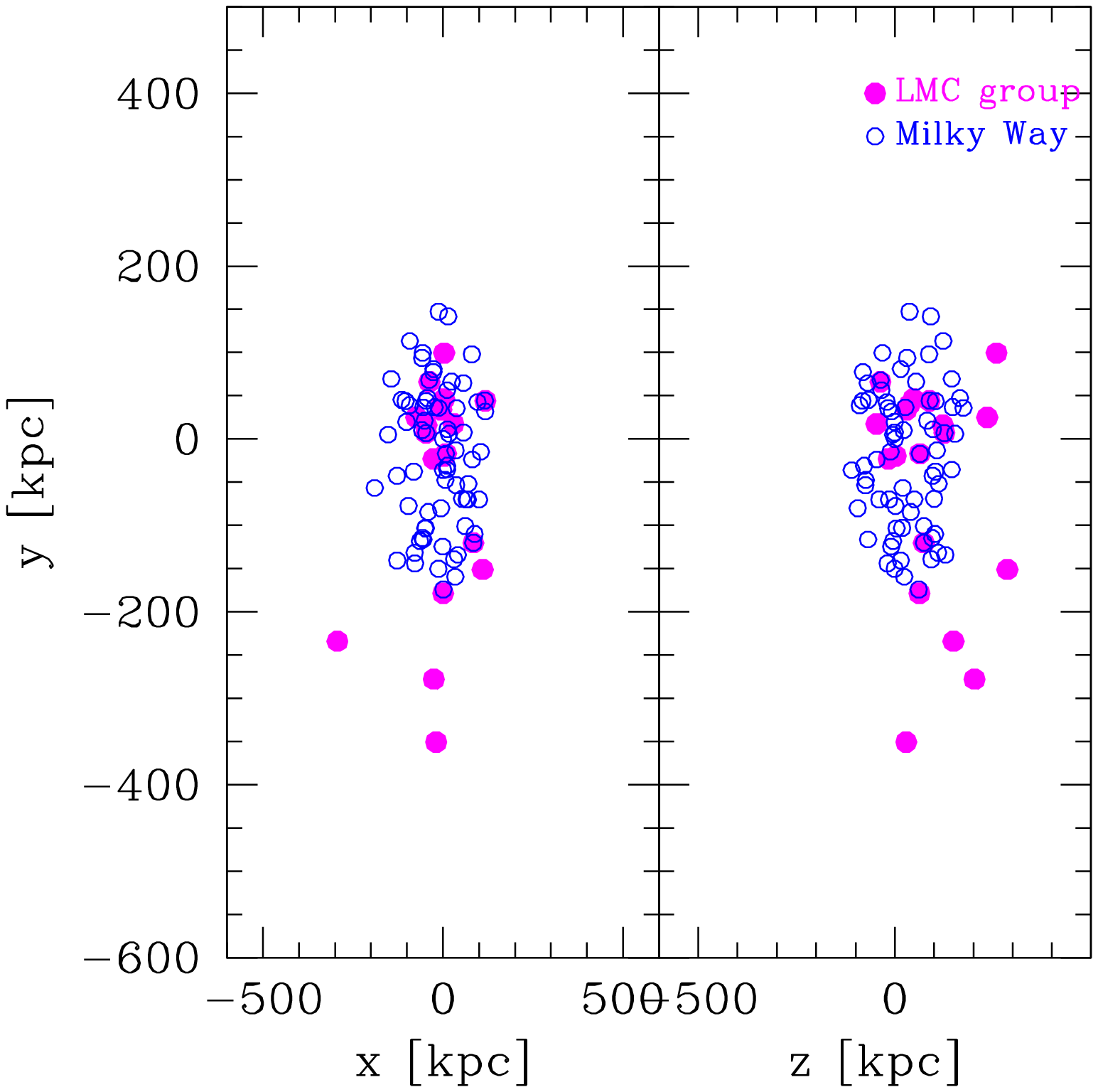} 
 \caption{The spatial distribution of satellites within the virial radius of the Milky Way (blue open circles) as compared
to the contributed subhalos from the break-up of the Magellanic Group at z=0 within 600 Kpc from the Milky Way center
(magenta filled circles).}
   \label{fig3}
\end{center}
\end{figure}

Figure 2 (left panel) shows the cumulative circular velocity
distribution function inferred for the dwarf associations, the
putative Magellanic Group (candidates listed previously), and the MW
satellite galaxies.  For each dwarf association we
assume the largest dwarf galaxy circular velocity of the group to be
the parent halo circular velocity.  Magnitudes of member galaxies are
converted to circular velocity assuming a Tully-Fisher relation in the
B band (see D'Onghia \& Lake (2008) for details).  
The MW data includes the newest dwarfs with a minimum $\sigma
= 3.3$ km s$^{-1}$ and a correction for incomplete sky coverage (Simon
\& Geha 2007).

Figure 2 shows that the nearby associations of dwarfs have a
cumulative circular velocity distribution function similar to the MW,
suggesting that such associations may be the progenitors of the
brightest dwarf satellites in the MW.  Thus, if these associations of
dwarfs are accreted into larger galaxies, they can populate the bright
end of the cumulative circular velocity distribution function of
satellites.  However, when normalized to the low mass of their parent,
they have a far greater number of dwarfs.

\section{Dwarfs in the LMC group can light up more efficiently}

In our interpretation, the mass of the LMC group is $\sim 4$\% of the
Milky Way, yet most of the dwarfs known a decade ago are associated
with it.  There is a similar overabundance of dwarfs in the dwarf
associations.  Here, we suggest that dwarf galaxies formed in LMC-like
groups will be luminous, while those that form by themselves in the
halos of larger systems will be dark.

It is generally assumed that galaxies with circular velocities $\sim
30$ km s$^{-1}$ blow out their gas.  When gas is blown out of a
subhalo, it eventually thermalizes to the virial temperature of the
parent halo, which is $2-5 \times 10^6$K for bright galaxies such as
the MW.  At this temperature, the cooling times are long enough that
there can be a considerable reservoir of hot gas and a subhalo with a
velocity scale of 10-30 km s$^{-1}$ will not reaccrete much gas, and it
will be dark.  However, in a small parent halo like the LMC, the
virial temperature is only 2x10$^5$K.  This is at the peak of the
cooling curve and the gas cools rapidly to $10^4K$.  The
low bulk motions in these halos might well permit reaccretion by some
of the subhalos producing luminous dwarf galaxies.  Note that our
picture is consistent with the new proper motion measurements from
Kallivayalil et al. 2006a,b and orbit models from Besla et al. (2007).
Prior to infall, the LMC group had a virial radius of $\sim$75 kpc and
a 3-D velocity dispersion of $\sim$ 100 km s$^{-1}$.  So, a thin plane
would still be very unusual and a wide range of kinematics is expected
for the disrupted satellites.

To investigate the plausibility of our model, we examined a catalog of
high resolution galaxies in a cosmologically simulated volume to
identify an analog to an LMC group with late infall into a MW galaxy.
We note in this specific simulation that the LMC group is tidally
disrupted before entering the virial radius of the MW, due to
the specific mass distribution of this case.  This could well be necessary
to prevent the merger of the LMC and SMC prior to accretion.
In Figure 2 (right panel), we display the cumulative peak circular
velocity distribution of the satellites contributed by the simulated
infalling group of dwarfs measured at z=0 within the virial radius of
the MW.  This is compared to the corresponding quantity for dwarfs
(filled squared symbols) in the MW which may have been part of an
accreted group: LMC, SMC, Sagittarius, Ursa Minor, Draco, Sextans and
Leo II.  In Figure 2, only
satellites that are accreted as part of the disrupted LMC group are
displayed, because those are the dwarf galaxies that light up in our
model.  The remainder of
the satellites that are not accreted in groups but
are within the virial radius of the present-day MW are assumed to
be dark.

We note that in this particular simulation, some satellites of the
disrupted group are outside the MW radius at z=0 and some are located
inside.  Figure 3 shows the spatial distribution of all the satellites
within the virial radius of the Milky Way (blue filled circles) as
compared to the subhalos of the disrupted Magellanic group at z=0
(magenta stars).  Despite the late infall, this particular group
appears very well mixed, however almost half of the surviving subhalos
of the group are at the present time located outside the virial radius
of the final Milky Way.  A few of them are in the outskirts of the
Milky Way. These subhalos may reproduce the special cases
like Tucana or Cetus that are located in low density regions of the Local
Group.

\section{Conclusion}

We assume a model where the LMC was the largest member of a group of
dwarf galaxies that was accreted into the MW halo.  Our picture addresses
several questions in galaxy formation: ({\it i}) It explains the
association of some dwarf galaxies in the Local group with the LMC-SMC
system.  ({\it ii}) It provides a mechanism to light up dwarf galaxies.
({\it iii}) It predicts that other isolated dwarfs will have companions.
The recent discovery of Leo V (Belokurov et al. 2008), a dwarf
spheroidal companion of Leo IV, and the nearby dwarf associations
supports our hypothesis.

E.D is grateful to Jacco van Loon and Joana Oliveira for organizing
an interesting meeting. She also would like to thank J. Gallagher,
G. Besla, K. Bekki, L. Hernquist,  N. Kallivayalil, C. Mastropietro 
for fruitful discussions.

\end{document}